\RequirePackage{fix-cm}
\RequirePackage{rotating}
\documentclass[smallextended]{svjour3}       % onecolumn (second format)
\smartqed  % flush right qed marks, e.g. at end of proof
\usepackage{graphicx}
\usepackage{url}
\usepackage{graphicx}
\usepackage{comment}
\usepackage{multirow}
\usepackage{color}
\usepackage{lscape}
\usepackage{longtable}
\usepackage{supertabular}
\usepackage{subfig}
\usepackage{graphicx}
\usepackage{natbib}
\usepackage[table]{xcolor}
\usepackage{pgfplots}
\usepackage[hidelinks,bookmarks=false]{hyperref}
\usepackage{enumitem}
\usepackage{booktabs}
\usepackage{amsmath,amssymb,amsfonts}
\usepackage{algorithmic}
\usepackage{textcomp}
\usepackage{listings}
\usepackage{pgfplotstable}
\def\BibTeX{{\rm B\kern-.05em{\sc i\kern-.025em b}\kern-.08em
    T\kern-.1667em\lower.7ex\hbox{E}\kern-.125emX}}
\pgfplotstableset{
    color cells/.style={
        col sep=comma,
        string type,
        postproc cell content/.code={%
                \pgfkeysalso{@cell content=\rule{0cm}{2.4ex}\cellcolor{black!##1}\pgfmathtruncatemacro\number{##1}\ifnum\number>50\color{white}\fi##1}%
                },
        columns/x/.style={
            column name={},
            postproc cell content/.code={}
        }
    }
}
\usepackage{multirow}
\usepackage{comment}
\definecolor{darkblue}{rgb}{0.0,0.0,0.6}

\newcommand\ok[1]{{\textcolor{black}{#1}}}

\newlength{\maxlen}

\begin{document}
\begin{sloppy}
\title{Wait For It:\\Identifying ``On-Hold'' Self-Admitted Technical Debt}
%\subtitle{sub}


\author{ Rungroj Maipradit         \and
        Christoph Treude         \and
        Hideaki Hata         \and
        Kenichi Matsumoto %etc.
}


% Hideaki Hata \at Nara Institute of Science and Technology, Japan \\ \email{hata@is.naist.jp} \and
% Kenichi Matsumoto \at Nara Institute of Science and Technology, Japan \\ \email{matumoto@is.naist.jp}
\institute{
Rungroj Maipradit \and Hideaki Hata \and Kenichi Matsumoto 
\at Nara Institute of Science and Technology \\ 
\email{maipradit.rungroj.mm6@is.naist.jp, hata@is.naist.jp, matumoto@is.naist.jp} \and
Christoph Treude \at University of Adelaide \\ \email{christoph.treude@adelaide.edu.au} \and
}

% \institute{F. Author \at
%               first address \\
%               Tel.: +123-45-678910\\
%               Fax: +123-45-678910\\
%               \email{fauthor@example.com}           %  \\
% %             \emph{Present address:} of F. Author  %  if needed
%               second address
% }

\date{Received: date / Accepted: date}

\maketitle

\begin{abstract}
Self-admitted technical debt refers to situations where a software developer knows that their current implementation is not optimal and indicates this using a source code comment. In this work, we hypothesize that it is possible to develop automated techniques to understand a subset of these comments in more detail, and to propose tool support that can help developers manage self-admitted technical debt more effectively. Based on a qualitative study of 335 comments indicating self-admitted technical debt, we first identify one particular class of debt amenable to automated management: ``on-hold'' self-admitted technical debt, i.e., debt which contains a condition to indicate that a developer is waiting for a certain event or an updated functionality having been implemented elsewhere. We then design and evaluate an automated classifier which can identify these ``on-hold'' instances with an \ok{area under the receiver operating characteristic curve (AUC)} of \ok{0.83} as well as detect the specific conditions that developers are waiting for.
Our work presents a first step towards automated tool support that is able to indicate when certain instances of self-admitted technical debt are ready to be addressed.
\end{abstract}

\section{Introduction}
\label{intro}
The metaphor of technical debt is used to describe the trade-off many software developers face when developing software: how to balance near-term value with long-term quality~\citep{Ernst2015}. Practitioners use the term technical debt as a synonym for ``shortcut for expediency''~\citep{McConnell2007} as well as to refer to bad code and inadequate refactoring~\citep{Kniberg2013}. Technical debt is widespread in the software domain and can cause increased software maintenance costs as well as decreased software quality~\citep{Lim2012}.

In many cases, developers know when they are about to cause technical debt, and they leave documentation to indicate its presence~\citep{Maldonado2017}. This documentation often comes in the form of source code comments, such as ``\texttt{TODO: This method is too complex, [let's] break it up}'' and ``\texttt{TODO no methods yet for getClassname}''.\footnote{Examples from ArgoUML and Apache Ant, respectively \citep{Maldonado2017}.} Previous work~\citep{Ichinose2016} has explored the use of visualization to support the discovery and removal of self-admitted technical debt, incorporating gamification mechanisms to motivate developers to contribute to the debt removal. Current research is largely focused on the detection and classification of self-admitted technical debt, \ok{but has spent less effort on approaches to address the debt automatically, likely because work on the detection and classification is still very recent.}

Previous work~\citep{Maldonado2017} has developed an approach based on natural language processing to automatically detect self-admitted technical debt comments and to classify them into either design or requirement debt. Self-admitted design debt encompasses comments that indicate problems with the design of the code while self-admitted requirement debt includes all comments that convey the opinion of a developer suggesting that the implementation of a requirement is incomplete. In general terms, design debt can be resolved by refactoring whereas requirement debt indicates the need for new code. 

\begin{figure*}
\centering
\begin{lstlisting}[language=Java]
// TODO the following code is copied from AbstractSimpleBeanDefinitionParser 
// it can be removed if ever the doParse() method is not final! 
// or the Spring bug http://jira.springframework.org/browse/SPR-4599 is resolved
\end{lstlisting}
\caption{Motivating Example, cf.~\url{https://github.com/apache/camel/blob/53177d55053a42f6fd33434895c60615713f4b78/components/camel-spring/src/main/java/org/apache/camel/spring/handler/BeanDefinitionParser.java}}
\label{fig:motivating}
\end{figure*}

In this work, we hypothesize that it is possible to use automated techniques based on natural language processing to understand a subset of the technical debt categories identified in previous work in more detail, and to propose tool support that can help developers manage self-admitted technical debt more effectively. We make three contributions:

\begin{itemize}
\item 
%A qualitative study of the removal 
\ok{A qualitative study on the removal} of self-admitted technical debt. To understand what kinds of technical debt could be addressed or managed automatically, we annotated a statistically representative sample of instances of self-admitted technical debt removal from the data set made available by the authors of previous work~\citep{MaldonadoICSME2017}. While the focus of our annotators was on the identification of instances of self-admitted technical debt that could be automatically addressed, as part of this annotation, we also performed a partial conceptual replication~\citep{Shull2008} of recent work by~\cite{Zampetti2018},\footnote{Note that~\cite{Zampetti2018} was published after we commenced this project, i.e., we do not use their data.} who found that a large percentage of self-admitted technical debt removals occur accidentally. We were able to confirm this finding: in 58\% of the cases in our sample, the self-admitted technical debt was not actually addressed, but the admission was simply removed. This finding is also in line with findings from~\cite{Bazrafshan2013} who reported a large number of accidental removals of cloned code. \ok{\cite{Zampetti2018} further reported that in removing self-admitted technical debt comments, developers tend to apply complex changes. Our work indirectly confirms this by finding that a majority of changes which address self-admitted technical debt could not easily be applied to similar debt in a different project.}
\item The definition of \textit{``on-hold'' self-admitted technical debt}. Our annotation revealed one particular class of self-admitted technical debt amenable to automated management: ``on-hold'' self-admitted technical debt. We define ``on-hold'' self-admitted technical debt as self-admitted technical debt which contains a condition to indicate that a developer is waiting for a certain event or an updated functionality having been implemented elsewhere. Figure~\ref{fig:motivating} shows an example of ``on-hold'' self-admitted technical debt from the Apache Camel project. The developer is waiting for an external event (the visibility of \texttt{doParse()} changing or an external bug being resolved) and the comment admitting the debt is therefore ``on hold''.
\item The design and evaluation of a classifier for self-admitted technical debt. Since software developers must keep track of many events and updates in any software ecosystem, it is unrealistic to assume that developers will be able to keep track of all self-admitted technical debt and of events that signal that certain self-admitted technical debt is now ready to be addressed. To support developers in managing self-admitted technical debt, we designed a classifier which can automatically identify those instances of self-admitted technical debt which are ``on hold'', and detect the specific events that developers are waiting for. Our classifier achieves an \ok{area under the receiver operating characteristic curve (AUC)} of \ok{0.83} for the identification, and
\ok{90}$\%$ of the specific conditions are detected correctly. %\todo{performance for detection}.
This is a first step towards automated tool support that can recommend to developers when certain instances of self-admitted technical debt are ready to be addressed.
\end{itemize}

The remainder of this paper is structured as follows: In Section~\ref{sec:method}, we present our research questions and the methods that we used for collecting and analyzing data for the qualitative study. The findings from this qualitative study are presented in Section~\ref{sec:qualfindings}. Section~\ref{sec:design} describes the design of our classifier to identify ``on-hold'' self-admitted technical debt, and we present the results of our evaluation of the classifier in Section~\ref{sec:evaluation}. Section~\ref{sec:implications} discusses the implications of this work, before Section~\ref{sec:threats} highlights the threats to validity and Section~\ref{sec:related} summarizes related work. Section~\ref{sec:conclusions} outlines the conclusions and highlights opportunities for future work. 

\section{Research Methodology}
\label{sec:method}

In this section, we detail our research questions as well as the methods for data collection and analysis used in our qualitative study. The methods for designing and evaluating our classifier are detailed in Sections~\ref{sec:design} and~\ref{sec:evaluation}. We also describe the data provided in our online appendix.

\subsection{Research Questions}

Our research questions focus on identifying how self-admitted technical debt is typically removed and whether the fixes applied to this debt could be applied to address similar debt in other projects. To guide our work, we first ask about the different kinds of self-admitted technical debt that can be found in our data (RQ1.1), whether the commits which remove the corresponding comments actually fix the debt (RQ1.2), and if so, what kind of fix has been applied (RQ1.3). To understand the removal in more detail, we also investigate whether the removal was the primary reason for the commit (RQ1.4), before investigating the subset of self-admitted technical debt that could be managed automatically (RQ1.5). Based on the definition of ``on-hold'' self-admitted technical debt which emerged from our qualitative study to answer these questions, we then investigate its prevalence (RQ1.6) and the accuracy of automated classifiers to identify this particular class of self-admitted technical debt (RQ2.1) and its specific sub-conditions (RQ2.2):

\begin{description}[labelsep=1em]
\item[RQ1] How do developers remove self-admitted technical debt?
    \begin{description}[labelsep=1em]
        \item[RQ1.1] What kinds of self-admitted technical debt do developers indicate?
        \item[RQ1.2] Do commits which remove the comments indicating self-admitted technical debt actually fix the debt?
        \item[RQ1.3] What kinds of fixes are applied to address self-admitted technical debt?
        \item[RQ1.4] Is the removal of self-admitted technical debt the primary reason for the commits which remove the corresponding comments?
        \item[RQ1.5] Could the fixes applied to address self-admitted technical debt be applied to address similar debt in other projects?
        \item[RQ1.6] How many of the comments indicating self-admitted technical debt contain a condition to specify that a developer is waiting for a certain event or an updated functionality having been implemented elsewhere?
    \end{description}
\item[RQ2] How accurately can \ok{our classifier} automatically identify ``on-hold'' self-admitted technical debt?
    \begin{description}[labelsep=1em]
        \item[RQ2.1] What is the best performance of \ok{our} classifier to automatically identify ``on-hold'' self-admitted technical debt?
        \item[RQ2.2] How well can \ok{our classifier} automatically identify the specific conditions in ``on-hold'' self-admitted technical debt?
    \end{description}
\end{description}

\subsection{Data Collection}

\begin{table}
    \centering
    \caption{Data set}
    \begin{tabular}{lrr}
    \toprule
    \textbf{project} & \textbf{SATD removal commits} & \textbf{sample} \\
    \midrule
    % Apache Camel & 1,050 & 130 \\
    % Apache Tomcat & 921 & 125 \\
    % Apache Hadoop & 374 & 52 \\
    % Gerrit Code Review & 146 & 19 \\
    % Apache Log4j & 108 & 9 \\
    % \midrule
    % Total & 2,599 & 335 \\
    Apache Camel & \ok{987} & 130 \\
    Apache Tomcat & \ok{910} & 125 \\
    Apache Hadoop & \ok{370} & 52 \\
    Gerrit Code Review & \ok{133} & 19 \\
    Apache Log4j & \ok{107} & 9 \\
    \midrule
    Total & \ok{2,507} & 335 \\
    \bottomrule
    \end{tabular}
    \label{tab:projects}
\end{table}

To obtain data on the removal of self-admitted technical debt, we used the online appendix of \cite{MaldonadoICSME2017} as a starting point. In their work, da Silva Maldonado et al.~conducted an empirical study on five open source projects to examine how self-admitted technical debt is removed, who removes it, for how long it lives in a project, and what activities lead to its removal. They make their data available in an online appendix\footnote{\url{http://das.encs.concordia.ca/uploads/2017/07/maldonado_icsme2017.zip}}, \ok{which contains 2,599 instances of a commit removing self-admitted technical debt. After removing duplicates, 2,507 instances remain.} The first two columns of Table~\ref{tab:projects} show the number of commits for each of the five projects available in this data set. Note that as a consequence of reusing this data set, we are implicitly also reusing \cite{MaldonadoICSME2017}'s definition of technical debt as well as their interpretation of what constitutes debt removal.

Based on this data set of commits which removed a comment indicating self-admitted technical debt (after removing duplicates), we created a statistically representative and random sample (confidence level 95\%, confidence interval \ok{4.98})\footnote{\url{https://www.surveysystem.com/sscalc.htm}} of 335 commits. The last column of Table~\ref{tab:projects} shows the number of commits from each project in our sample.

\subsection{Data Analysis}

To answer our first research question ``How do developers remove self-admitted technical debt?'' and its sub-questions, we performed a qualitative study on the sample of 335 commits which had removed self-admitted technical debt according to the data provided by~\cite{MaldonadoICSME2017}. 

\begin{sidewaystable}
        \centering
        \caption{Qualitative annotation schema}
        \begin{tabular}{p{6cm}lp{9.5cm}}
        \toprule
        \textbf{question} & \textbf{answers} & \textbf{motivation} \\
        \midrule
        Does the comment represent self-admitted technical debt? & yes/no & Observation that some comments that \cite{MaldonadoICSME2017} had automatically identified as self-admitted technical debt did not actually constitute debt\\
    %    Where is the comment describing the self-admitted technical debt? & main/test & Observation that debt found in test files seemed to be handled differently compared to debt outside of test files\\
        \textbf{RQ1.1} What kind of self-admitted technical debt was it? & open & Distinguishing different kinds of self-admitted technical debt, with the ultimate goal of identifying ones that can be addressed automatically\\
        \textbf{RQ1.2} Did the commit fix the self-admitted technical debt? & yes/no & Observation that commits which remove self-admitted technical debt do not necessarily fix the debt, as also found by \cite{Zampetti2018}\\
        \textbf{RQ1.3} What kind of fix was it? & open & Distinguishing different kinds of fixes for self-admitted technical debt, to study whether fixes could be applied automatically\\
        \textbf{RQ1.4} Was removing the self-admitted technical debt the primary reason for the commit? & yes/no & Observation that even for those commits which addressed self-admitted technical debt, this was not necessarily their main purpose\\
        \textbf{RQ1.5} Could the same fix be applied to similar self-admitted technical debt in a different project? & possibly/no & Identifying fixes that could potentially be applied automatically\\
        \textbf{RQ1.6} Does the self-admitted technical debt comment include a condition? & yes/no & Exploring the phenomenon of ``on-hold'' self-admitted technical debt---which emerged from answering the previous question---in more detail\\
        \bottomrule
        \end{tabular}
        \label{tab:annotationquestions}
\end{sidewaystable}

In the first step, the second and third author of this paper independently analyzed twenty commits from the sample to determine appropriate questions to be asked during the qualitative study, aiming to obtain insights into how developers remove self-admitted technical debt and to identify the kinds of debt that could be addressed or managed automatically. After several iterations and meetings, the second and third author agreed on seven questions that should be answered for each of the 335 commits during the qualitative study. These questions along with their motivation and answer ranges are shown in Table~\ref{tab:annotationquestions}.

The first author annotated all 335 commits following this annotation schema, and the second and third author annotated 50\% of the data each, ensuring that each commit was annotated according to all seven questions by two researchers. Note that not all questions applied to all commits. For example, all instances which we classified as not representing self-admitted technical debt were not considered for future questions, and all commits which we classified as not fixing self-admitted technical debt were not considered for questions such as ``Could the same fix be applied to similar Self-Admitted Technical Debt in a different project?''.

After the annotation, the first three authors conducted multiple meetings in which they determined consistent coding schemes for the two questions which allowed for open answers and collaboratively resolved all disagreements in the annotation until reaching consensus on all ratings. We report the initial agreement for each question before the resolution of disagreements as part of our findings in the next section.\footnote{We calculated kappa values using \url{https://www.graphpad.com/quickcalcs/kappa1/}.}

\subsection{Online Appendix}

Our online appendix contains descriptive information on the 335 commits which were labeled as removing self-admitted technical debt according to~\cite{MaldonadoICSME2017} along with our qualitative annotations in response to the seven questions. The appendix is available at \url{https://tinyurl.com/onholddebt}.

\section{Qualitative Findings}
\label{sec:qualfindings}

In this section, we describe the findings derived from our qualitative study, separately for each sub-question of RQ1.

\subsection{Initial Analysis}

\begin{figure}
\centering
\begin{footnotesize}
\begin{tikzpicture}
\begin{axis}[
    align=center,
    title = {Does the comment represent\\Self-Admitted Technical Debt?},
    y=0.5cm,
    x=0.024cm,
    enlarge y limits={abs=0.25cm},
    symbolic y coords={N/A, no, yes},
    axis line style={opacity=0},
    major tick style={draw=none},
    ytick=data,
    xmin = 0,
    nodes near coords,
    nodes near coords align={horizontal},
    point meta=rawx
]
\addplot[xbar,fill=gray,draw=none] coordinates {
    (286,yes)
    (19,no)
    (30,N/A)
};
\end{axis}
\end{tikzpicture}
\end{footnotesize}
\caption{Distribution of answers to ``Does the comment represent Self-Admitted Technical Debt?''. Initial agreement among the annotators before resolving disagreements: weighted kappa $\kappa=0.821$ across 335 comments, i.e., ``almost perfect'' agreement~\citep{viera2005understanding}.}
\label{fig:commentisdebt}
\end{figure}
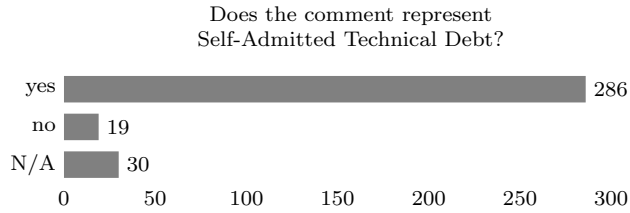

As shown in Figure~\ref{fig:commentisdebt}, we found that not all commits which were automatically classified as removing self-admitted technical debt by the work of~\cite{MaldonadoICSME2017} actually removed a comment indicating debt. In some cases (9\%)---indicated as N/A in Figure~\ref{fig:commentisdebt}---the comment was not removed but only edited, and in other cases (6\%), the comment had been incorrectly tagged as self-admitted technical debt, e.g., in the case of ``\texttt{It is always a good idea to call this method when exiting an application}''.

\subsection{RQ1.1 What kinds of self-admitted technical debt do developers indicate?}

\begin{figure}
\centering
\begin{footnotesize}
\begin{tikzpicture}
\begin{axis}[
    align=center,
    title = {\textbf{RQ1.1} What kind of Self-Admitted\\Technical Debt was it?},
    y=0.5cm,
    x=0.024cm,
    enlarge y limits={abs=0.25cm},
    symbolic y coords={N/A,other,explanation,bug,wait,workaround,clarification request,refactoring needed,functionality needed},
    axis line style={opacity=0},
    major tick style={draw=none},
    ytick=data,
    xmin = 0,
    nodes near coords,
    nodes near coords align={horizontal},
    point meta=rawx
]
\addplot[xbar,fill=gray,draw=none] coordinates {
    (126,functionality needed)
    (49,refactoring needed)
    (43,clarification request)
    (24,workaround)
    (13,wait)
    (12,bug)
    (5,explanation)
    (14,other)
%    (49,N/A)
};
\end{axis}
\end{tikzpicture}
\end{footnotesize}
\caption{Distribution of answers to ``What kind of Self-Admitted Technical Debt was it?''. Initial agreement among the annotators before consolidating the coding schema: 45.45\% across 286 comments.}
\label{fig:debtkinds}
\end{figure}
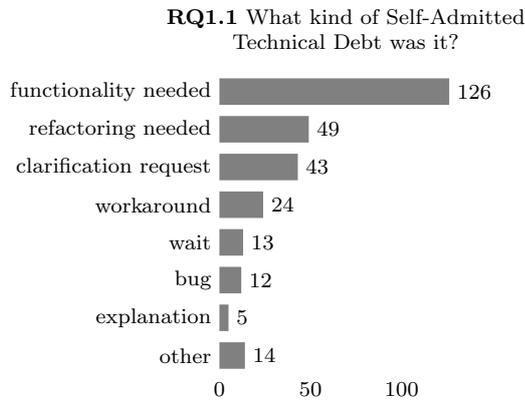

Our first research question explores the different kinds of self-admitted technical debt found in our sample. Figure~\ref{fig:debtkinds} shows the final result of our coding after consolidating the coding schema. The two most common kinds of debt in our sample are ``functionality needed'' (44\%) and ``refactoring needed'' (17\%). An example for the former is the comment ``\texttt{TODO handle known multi-value headers}'' while ``\texttt{XXX move message resources in this package}'' is an example for the latter. We also identified a number of clarification requests (15\%), such as ``\texttt{TODO: why not use millis instead of nano?}''. We coded self-admitted technical debt comments that explicitly stated that they were temporary as workaround (8\%), e.g., ``\texttt{TODO this should subtract resource just assigned TEMPROARY}''. We identified some comments which indicated that the developer was waiting for something (5\%), such as ``\texttt{TODO remove these methods if/when they are available in the base class!!!}''. We will focus our discussion on these comments in the later parts of this paper. Finally, some comments which indicated technical debt describe bugs (4\%, e.g., ``\texttt{TODO this causes errors on shutdown...}'') or focus on explaining the code (2\%, e.g., ``\texttt{some OS such as Windows can have problem doing rename IO operations so we may need to retry a couple of times to let it work}''). Note that for this annotation, we assigned exactly one code to each comment.

\ok{Previous classifications of self-admitted technical debt focused less on the actions required to remove the debt and more on what part of the software development lifecycle a debt item can be assigned to. For example, the categorisation of \cite{7332619} revealed five categories (design, defect, documentation, requirement, and test), and the categorisation of Bavota and Russo~\cite{7832911} revealed the same five categories plus a sixth category called ``code''. In comparison, guided by our ultimate goal of identifying certain kinds of self-admitted technical debt which can be fixed automatically, our categorisation focuses more on what needs to be done in order to fix the debt, leading to categories such as ``functionality needed'' or ``refactoring needed''.}

\subsection{RQ1.2 Do commits which remove the comments indicating self-admitted technical debt actually fix the debt?}

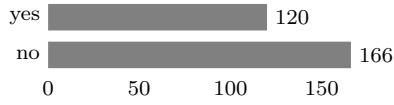
\begin{figure}
\centering
\begin{footnotesize}
\begin{tikzpicture}
\begin{axis}[
    align=center,
    title = {\textbf{RQ1.2} Did the commit fix the Self-Admitted\\Technical Debt?},
    y=0.5cm,
    x=0.024cm,
    enlarge y limits={abs=0.25cm},
    symbolic y coords={N/A, no, yes},
    axis line style={opacity=0},
    major tick style={draw=none},
    ytick=data,
    xmin = 0,
    nodes near coords,
    nodes near coords align={horizontal},
    point meta=rawx
]
\addplot[xbar,fill=gray,draw=none] coordinates {
    (120,yes)
    (166,no)
%    (49,N/A)
};
\end{axis}
\end{tikzpicture}
\end{footnotesize}
\caption{Distribution of answers to ``Did the commit fix the Self-Admitted Technical Debt?''. Initial agreement among the annotators before resolving disagreements: kappa $\kappa=0.734$ across 286 comments, i.e., ``substantial'' agreement~\citep{viera2005understanding}.}
\label{fig:debtremove}
\end{figure}

For the majority of commits (58\%) which removed the comment indicating technical debt, the commit did not actually fix the problem described in the comment, see Figure~\ref{fig:debtremove}. Instead, these commits often removed the comment along with the surrounding code. These findings are in line with recent work by~\cite{Zampetti2018} who found that between 20\% and 50\% of self-admitted technical debt is accidentally removed while entire classes or methods are dropped.

\subsection{RQ1.3 What kinds of fixes are applied to address self-admitted technical debt?}

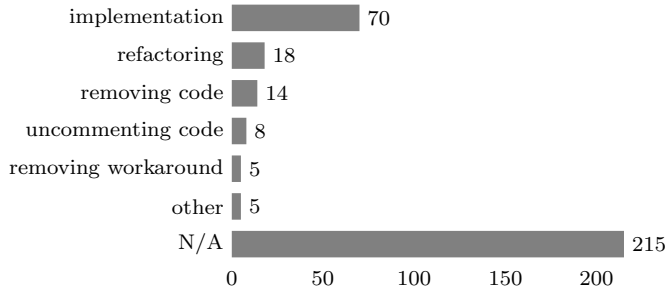
\begin{figure}
\centering
\begin{footnotesize}
\begin{tikzpicture}
\begin{axis}[
    align=center,
    title = {\textbf{RQ1.3} What kind of fix was it?},
    y=0.5cm,
    x=0.024cm,
    enlarge y limits={abs=0.25cm},
    symbolic y coords={N/A,other,removing workaround,uncommenting code,removing code,refactoring,implementation},
    axis line style={opacity=0},
    major tick style={draw=none},
    ytick=data,
    xmin = 0,
    nodes near coords,
    nodes near coords align={horizontal},
    point meta=rawx
]
\addplot[xbar,fill=gray,draw=none] coordinates {
    (70,implementation)
    (18,refactoring)
    (14,removing code)
    (8,uncommenting code)
    (5,removing workaround)
    (5,other)
    (215,N/A)
};
\end{axis}
\end{tikzpicture}
\end{footnotesize}
\caption{Distribution of answers to ``What kind of fix was it?''. Initial agreement among the annotators before consolidating the coding schema: 84.17\% across 120 comments.}
\label{fig:fixkinds}
\end{figure}

In the cases where the commit fixed the self-admitted technical debt, we also coded the kind of fix that was applied. Figure~\ref{fig:fixkinds} show the results of this coding: Debt was either fixed by implementing new code (58\%), by refactoring existing code (15\%), by removing code (12\%), by uncommenting code that had been previously commented out (7\%), or by removing a workaround (4\%). Note that we used the commit message and/or related issue discussions to determine whether a change was meant to remove a workaround or was truly a refactoring. \ok{Other cases, such as uncommenting code, were easy to decide.}
\ok{
 In the 215 cases where the commit does not fix the self-admitted technical debt, 30 commits do not remove the self-admitted technical debt comments or are tagged incorrectly, 19 comments do not represent self-admitted technical debt, and 166 commits do not fix self-admitted technical debt.
}
\ok{
Our categorisation of the different kinds of fixes is at a slightly more coarse-granular level compared to that presented by \cite{Zampetti2018} who identified five categories (add/remove method calls, add/remove conditionals, add/remove try-catch, modify method signature, and modify return) in addition to ``other''. In their categorisation, ``other'' accounts for 44\% (339/779) of all instances. In comparison, our categorisation is less fine-grained, but contains fewer ``other'' cases.
}
% \begin{table*}
\begin{sidewaystable}
\caption{Types of Self-Admitted Technical Debt and the Corresponding Fixes. Each row represents a type of self-admitted technical debt, and each column represents a type of fix. The sum of each row and column indicates the overall numbers for the corresponding codes, respectively.}
\centering
\pgfplotstabletypeset[color cells]{
x,implementation,refactoring,removing code,uncommenting code,removing workaround,other,not fixed
functionality needed,56,1,0,0,0,0,69
refactoring needed,2,16,1,0,0,1,29
clarification request,5,0,4,0,0,1,33
workaround,2,0,2,3,5,0,12
wait,2,0,2,3,0,0,6
bug,2,0,1,2,0,1,6
explanation,0,0,0,0,0,0,5
other,1,1,4,0,0,2,6
}
\label{tab:debtandfix}
% \end{table*}
\end{sidewaystable}

Table~\ref{tab:debtandfix} shows the relationship between the two coding schemes that emerged from our qualitative data analysis: one for the kinds of technical debt indicated in developer comments, and one for the kinds of fixes applied to this debt. Unsurprisingly, many instances where new functionality was needed were addressed by the implementation of said functionality, and cases where refactoring was needed were addressed by refactoring. Interestingly, all comments of developers explaining technical debt were removed without addressing the debt\ok{. An example is the self-admitted technical debt comment ``\texttt{some OS such as Windows can have problem doing delete IO operations so we may need to retry a couple of times to let it work}'' in the Apache Camel project which was removed in commit f10f55e\footnote{\url{https://github.com/apache/camel/commit/f10f55e38945686827dc249703b1606682657a62}} together with the surrounding source code. We hypothesise that in some cases, developers decide to replace code which requires an explanation with simpler code. More work will have to be conducted to test this hypothesis.} Waits could sometimes be addressed by uncommenting code that had been written in anticipation of the fix. A large number of comments indicating debt were not addressed---for example, out of 43 comments which we coded as clarification request, 33 (77\%) were ``resolved'' by simply deleting the comment \ok{(e.g., the comment ``\texttt{TODO why zero?}'' was removed from the Apache Camel source code in commit 3d8f4e9\footnote{\url{https://github.com/apache/camel/commit/3d8f4e9d68253269b4f5cf7e3cfea4553b46d74f}} without further explanation}. Note that in cases where more than one of our codes could apply, we noted the most prominent one. \ok{This could for example occur in cases of long comments which were used to communicate different concerns. In such rare cases, we applied the code for the longest section of the comment.} This explains the small number of inconsistencies, e.g., a ``functionality needed'' debt fixed by a ``refactoring''.

\subsection{RQ1.4 Is the removal of self-admitted technical debt the primary reason for the commits which remove the corresponding comments?}

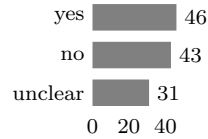
\begin{figure}
\centering
\begin{footnotesize}
\begin{tikzpicture}
\begin{axis}[
    align=center,
    title = {\textbf{RQ1.4} Was removing the Self-Admitted Technical Debt\\the primary reason for the commit?},
    y=0.5cm,
    x=0.024cm,
    enlarge y limits={abs=0.25cm},
    symbolic y coords={N/A, unclear, no, yes},
    axis line style={opacity=0},
    major tick style={draw=none},
    ytick=data,
    xmin = 0,
    nodes near coords,
    nodes near coords align={horizontal},
    point meta=rawx
]
\addplot[xbar,fill=gray,draw=none] coordinates {
    (46,yes)
    (43,no)
    (31,unclear)
%    (215,N/A)
};
\end{axis}
\end{tikzpicture}
\end{footnotesize}
\caption{Distribution of answers to ``Was removing the Self-Admitted Technical Debt the primary reason for the commit?''. Agreement among the annotators: weighted kappa $\kappa=0.696$ across 120 comments, i.e., ``substantial'' agreement~\citep{viera2005understanding}.}
\label{fig:primaryreason}
\end{figure}

The removal of technical debt was often not the primary reason for commits which removed self-admitted debt, see Figure~\ref{fig:primaryreason}. %This is in line with findings reported by~\cite{Zampetti2018} who found that only 8\% of the technical debt removal is acknowledged in commit messages. 
We did not attempt to resolve disagreements between annotators for this question as the concept of ``primary reason'' can be ambiguous. Instead, instances where annotators disagreed are shown as ``unclear'' in Figure~\ref{fig:primaryreason}.

An example of a commit which removed self-admitted technical debt even though it was not the main purpose of the commit is Apache Camel commit f47adf.\footnote{\url{https://github.com/apache/camel/commit/f47adf75510ef71a5b4071e8c77af7abb9c07dc9}} The commit removed the following comment: ``\texttt{TODO: Support ordering of interceptors}'', but this was part of a much larger refactoring as described in the commit message: ``\texttt{Overhaul of JMX}''. On the other hand, the commit message of commit 88ca35\footnote{\url{https://github.com/apache/camel/commit/88ca359343c3a96786d435985f46841eeffcfb6e}} from the same project ``\texttt{Added onException support to DefaultErrorHandler}'' is very similar to the self-admitted technical debt comment that was removed in this commit ``\texttt{TODO: in the future support onException}'', which suggests that removing the debt was the primary reason for this commit.

\subsection{RQ1.5 Could the fixes applied to address self-admitted technical debt be applied to address similar debt in other projects?}

\begin{figure}
\centering
\begin{footnotesize}
\begin{tikzpicture}
\begin{axis}[
    align=center,
    title = {\textbf{RQ1.5} Could the same fix be applied to similar\\Self-Admitted Technical Debt in a different project?},
    y=0.5cm,
    x=0.024cm,
    enlarge y limits={abs=0.25cm},
    symbolic y coords={N/A, no, possibly},
    axis line style={opacity=0},
    major tick style={draw=none},
    ytick=data,
    xmin = 0,
    nodes near coords,
    nodes near coords align={horizontal},
    point meta=rawx
]
\addplot[xbar,fill=gray,draw=none] coordinates {
    (40,possibly)
    (80,no)
%    (215,N/A)
};
\end{axis}
\end{tikzpicture}
\end{footnotesize}
\caption{Distribution of answers to ``Could the same fix be applied to similar Self-Admitted Technical Debt in a different project?''. Agreement among the annotators: kappa $\kappa=0.541$ across 120 comments, i.e., ``moderate'' agreement~\citep{viera2005understanding}.}
\label{fig:autofix}
\end{figure}
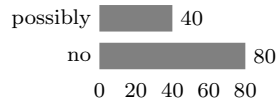

\ok{We annotated the 120 self-admitted technical debt comments which had been fixed by a commit in terms of whether the fix applied in this commit could be applied in a similar context in a different project. While this annotation was subjective to some extent---as also indicated by our kappa agreement of $0.541$ which was the lowest across all questions we answered about the self-admitted technical debt comments---we used our intuition about whether we could envision tool support to address a comment automatically. We used our experience of conducting research on automated tool support for source code manipulation for this step.}

We identified two kinds of self-admitted technical debt that could possibly be handled automatically. The first kind are comments which are fairly specific, e.g., ``\texttt{TODO gotta catch RejectedExecutionException and properly handle it}''. Automated tool support could be built to at least catch the exception based on this description. The second kind are comments which indicate that a developer is waiting for something, which we will discuss further in the next subsection. Figure~\ref{fig:autofix} shows the ratio of fixes that could possibly be automated and applied in other settings, which is one third of all fixes. Note that we counted all those comments as ``possibly'' that were rated as ``possibly'' by at least one annotator. This finding supports~\cite{Zampetti2018} who found that most changes addressing self-admitted technical debt require complex source code changes. The primary goal of investigating this research question was the identification of types of self-admitted technical debt likely amenable to being fixed automatically.

\subsection{RQ1.6 How many of the comments indicating self-admitted technical debt contain a condition to specify that a developer is waiting for a certain event or an updated functionality having been implemented elsewhere?}

\begin{figure}
\centering
\begin{footnotesize}
\begin{tikzpicture}
\begin{axis}[
    align=center,
    title = {\textbf{RQ1.6} Does the Self-Admitted Technical Debt\\comment include a condition?},
    y=0.5cm,
    x=0.024cm,
    enlarge y limits={abs=0.25cm},
    symbolic y coords={N/A, no, yes},
    axis line style={opacity=0},
    major tick style={draw=none},
    ytick=data,
    xmin = 0,
    nodes near coords,
    nodes near coords align={horizontal},
    point meta=rawx
]
\addplot[xbar,fill=gray,draw=none] coordinates {
    (27,yes)
    (259,no)
%    (49,N/A)
};
\end{axis}
\end{tikzpicture}
\end{footnotesize}
\caption{Distribution of answers to ``Does the Self-Admitted Technical Debt comment include a condition?''. Initial agreement among the annotators before resolving disagreements:  weighted kappa $\kappa=0.618$ across 286 comments, i.e., ``substantial'' agreement~\citep{viera2005understanding}.}
\label{fig:condition}
\end{figure}
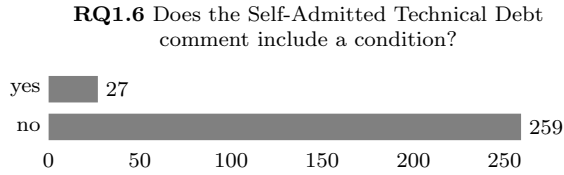

\begin{table}
    \centering
    \caption{Example of self-admitted technical debt on ``on-hold'' and ``wait''}
    \begin{tabular}{lr}
    \toprule
    \textbf{Example of SATD} & \textbf{Category / On-hold or not}\\
    \midrule
    \ok{// TODO change to file when this is ready} & \ok{wait / non on-hold}\\
    \ok{// FIXME: Code to be used in case} & \ok{wait / non on-hold}\\
    \ok{route replacement is needed} & \\
    \ok{// TODO: is needed when we add} & \ok{add functionality / on-hold}\\
    \ok{support for when predicate} & \\
    \ok{// TODO: Camel 2.9/3.0 consider} & \ok{refactor / on-hold}\\
    \ok{moving to org.apache.camel} & \\
    \bottomrule
    \end{tabular}
    \label{tab:onhold_wait}
\end{table}

A theme that emerged from answering the previous research question is the concept of self-admitted technical debt comments which include a condition to indicate that a developer is waiting for a certain event or an updated functionality having been implemented elsewhere. \ok{Since no other obvious class of self-admitted technical debt emerged which seemed amenable to automated tool support, we focus on this kind of self-admitted technical debt for building a classifier (see next section).} We refer to this kind of debt as ``on-hold'' self-admitted technical debt---the comment is ``on hold'' until the condition is met (see Figure~\ref{fig:motivating} for examples). In our sample, we identified 27 such comments, see Figure~\ref{fig:condition}. These comments are also related to the ``wait'' category shown in Figure~\ref{fig:debtkinds}, but not necessarily identical since the question addressed by Figure~\ref{fig:debtkinds} did not explicitly ask about conditions.
\ok{Table~\ref{tab:onhold_wait} shows examples of ``on-hold'' comments and those classified in the ``wait'' category.}

%\subsection{Summary}

%In summary, we were able to confirm the findings of previous work which indicate that the main categories of technical debt are related to requirements and design~\citep{Maldonado2017}. In the majority of cases, a commit which removes a comment admitting technical debt does not actually fix the debt, and even in the remaining cases, fixing the debt is often not the primary reason for the commit. When self-admitted technical debt is fixed, this is usually done through the implementation of new functionality, but we also identified cases where debt could be addressed by either removing or uncommenting code. We identified a particular sub-class of self-admitted technical debt comments which might be amenable to automated tool support, i.e., ``on-hold'' self-admitted technical debt. Next, we will describe the classifier we built to detect this subclass of self-admitted technical debt automatically.

\section{Classifier Design}
\label{sec:design}

\begin{figure*}[!t]
\centering
\includegraphics[width=\linewidth]{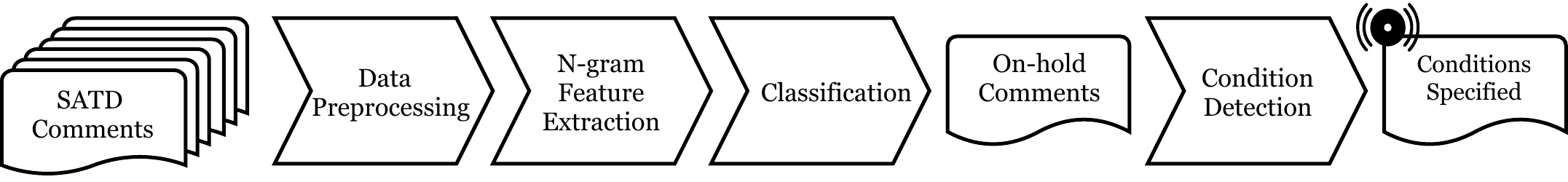}
\caption{Classification overview.}
\label{fig:overview}
\end{figure*}

Figure~\ref{fig:overview} shows the overview of our classifier for ``on-hold'' self-admitted technical debt identification and the detection of the specific conditions that developers are waiting for. Given self-admitted technical debt comments, data preprocessing and n-gram feature extraction are applied before classifying them into ``on-hold'' or not. Within identified ``on-hold'' self-admitted technical debt comments, specific conditions are detected.

\subsection{Data Preprocessing}

Three preprocessing steps are applied, namely, term abstraction, lemmatization, and special character removal.

\begin{table}[!t]
\renewcommand{\arraystretch}{1.3}
\centering
\caption{Regular expressions for term abstraction}
\label{tab:reg}
\begin{tabular}{lp{6cm}}
\toprule
\textbf{abstraction} & \textbf{pattern} \\
\midrule
@abstractdate & \texttt{(0[1-9]|[12]\textbackslash d|3[01]).(0[1-9]|1[0-2])}\\
& \texttt{.([12]\textbackslash d{3})} \\
 & \# year.month.date, e.g., 21.02.2011 \\
 & \texttt{(0[1-9]|[12]\textbackslash d|3[01])\textbackslash/(0[1-9]|1[0-2])}\\
 & \texttt{(\textbackslash/([12]\textbackslash d{3}))*} \\
 & \# day/month(/year), e.g., 25/05, 22/05/2012 \\
 & \texttt{((([0-9])|([0-2][0-9])|([3][0-1]))} \\
 & \texttt{(Jan|Feb|Mar|Apr|May|Jun|Jul|Aug|Sep|}\\
 & \texttt{Oct|Nov|Dec)\textbackslash w+ \textbackslash d{4}} \\
 & \# day month year, e.g., 23 June 2013 \\
 & \texttt{\textbackslash d+-\textbackslash d+-\textbackslash d+ \textbackslash d+:\textbackslash d+:\textbackslash d+ [-|+]\textbackslash d+} \\
 & \# year-month-day timestamp, e.g., 2006-03-06 23:16:24 +0100 \\
\midrule
@abstractversion & \texttt{[0-9]\{1,2\}\textbackslash.[0-9]\{1,2\}([+-]|\textbackslash.[0-9]\{1,3\}|} \\ &
\texttt{\textbackslash.[A-Za-z]\{1,2\})*(\textunderscore[0-9]\{1,3\})*} \\
 & \# release version, e.g., 1.9.3, 4.0, 8.0.x, 1.0.12\textunderscore25 \\
% \texttt{\textbackslash(\textbackslash d+\textbackslash.\textbackslash d+\textbackslash.\textbackslash d+\textbackslash)} \\
% & \texttt{[\textbackslash s-]\textbackslash d+\textbackslash.[\textbackslash d\textbackslash.\textbackslash w\textbackslash+]+} \\
% & \# release version, e.g., 4.0, 8.0.x \\
% & \texttt{([A-Za-z]+)(\textbackslash d+\textbackslash.\textbackslash d+\textbackslash.\textbackslash d+)} \\
% & \# release version, e.g., JDK1.1.8 \\
\midrule
% @abstractbugid & \texttt{([A-Za-z]+)-(\textbackslash d+)} \\
@abstractbugid & \texttt{abstractproduct[ |-]*\textbackslash d+} \\
& \# bug id, e.g., jetty-9.3 \\
\midrule
@abstracturl & \texttt{https?:\textbackslash/\textbackslash/(www\textbackslash.)?[-a-zA-Z0-9@:\%.\_\textbackslash}\\
& \texttt{+\textasciitilde\#=]\{2,256\}\textbackslash.[a-z]\{2,6\}\textbackslash b} \\
& \texttt{([-a-zA-Z0-9@:\%\_\textbackslash+.\textasciitilde\#?\&//=]*)}\\
& \# url \\
% \midrule
% @abstractproduct & \texttt{(tomcat|jdk|hdfs|hadoop|servlet|}\\
% & \texttt{cxf|camel|atmosphere|mapreduce|} \\
% & \texttt{yarn|hazelcast|catalina|hibernate)} \\
% & \# product name\\
\bottomrule
\end{tabular}
\end{table}

\begin{figure*}[!t]
\centering
\includegraphics[width=.8\linewidth]{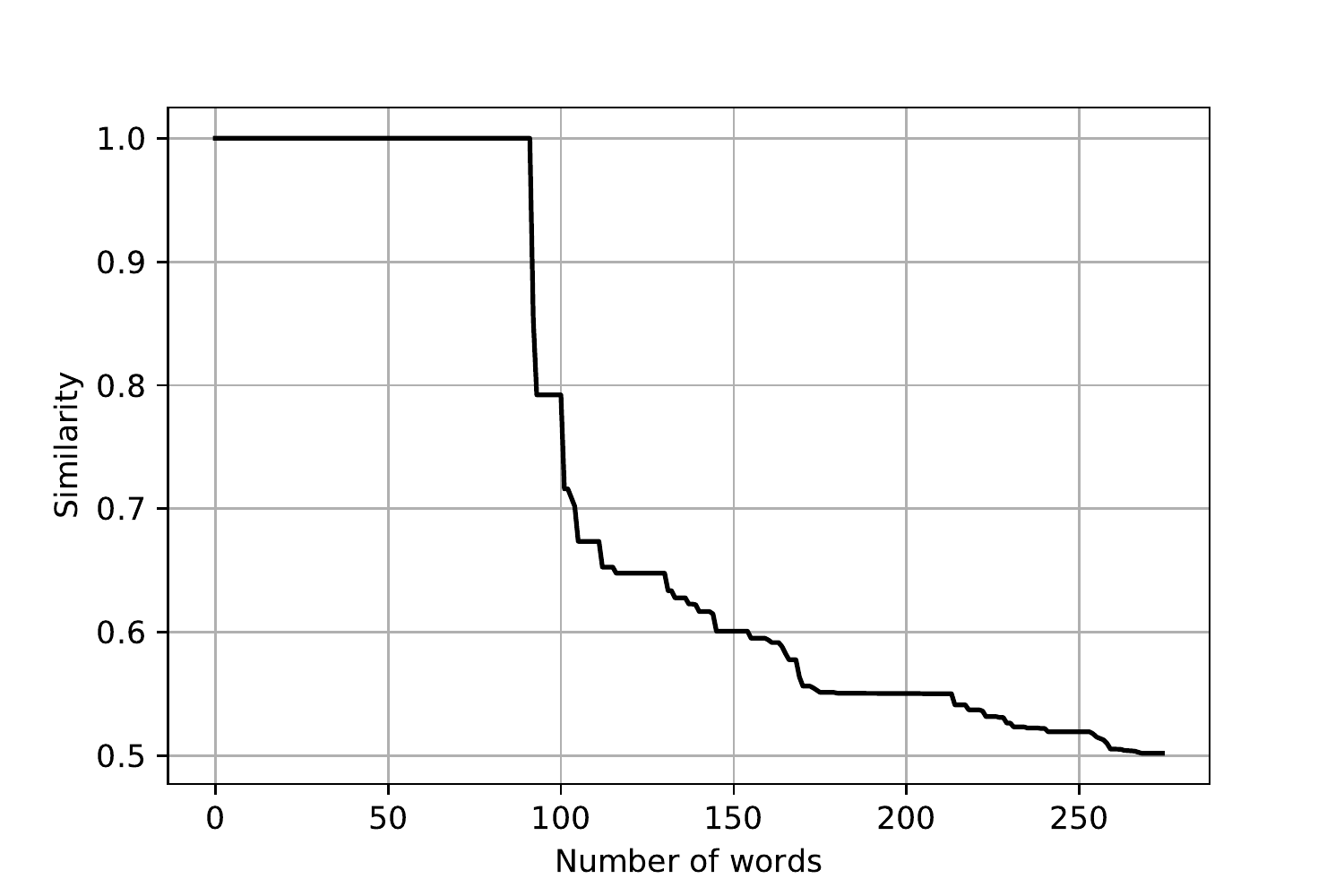}
\caption{Similarity between project names and words.}
\label{fig:similarity}
\end{figure*}

%\subsubsection{Term Abstraction}
%\noindent
\paragraph{Term Abstraction.}

\ok{Term Abstraction. Similar to a previous text classification study~\citep{2018arXiv180206997A}, we perform abstraction as a preprocessing step. The previous study~\citep{2018arXiv180206997A} abstracted keywords from GitHub README files. Their abstraction included mail-to links, hyperlinks, code blocks, images,
    and numbers. We also apply abstraction for hyperlinks (URLs), however, we do not apply the others because images, mail-to links, and code blocks do not usually appear in comments.   
    Instead, we introduce four kinds of abstraction which are related to ``on-hold'' conditions.}
We target the following terms: \textit{date expression}, \textit{version}, \textit{bug id}, \textit{URL}, and \textit{product name}. 
Each term is abstracted into a string: \texttt{@abstractdate}, \texttt{@abstractversion}, \texttt{@abstractbugid}, \texttt{@abstracturl}, and \texttt{@abstractproduct}.
Table~\ref{tab:reg} shows the regular expressions we use to detect \texttt{@abstractdate}, \texttt{@abstractversion}, \texttt{@abstractbugid}, and \texttt{@abstracturl}.

For abstracting product names for \texttt{@abstractproduct}, we try finding semantically similar words to the project names and their sub-project names in our data set, i.e.,
%In the case of \textit{@abstractproduct} we apply by finding the similarity between each word in comments and project name and their sub project in datasets which is,
% Apache, Camel, Tomcat, Log4j, Hadoop, Yarn, Mapreduce, and Hdfs.
\ok{Apache, Camel, Tomcat, Hadoop, Gerrit, Log4j, Yarn, Mapreduce, Hdfs, Ant, Jmeter, Argouml, Columba, Emf, Hibernate, Distribution, Jedit, Jfreechart, Jruby, and Squirrel.}
Figure~\ref{fig:similarity} shows the similarity between each word in comments and project name and their related project using Spacy~\citep{spacy2}.\footnote{Spacy has recently been found to achieve a higher accuracy when applied to software-related text compared to other libraries~\citep{AlOmran2017}.}
According to the result, the similarity score drops drastically from 1.0---therefore, %we set the threshold at 0.9.
we consider words with similarity 1.0 as project names.
%After applying the threshold,
We obtained a set of words including 
% Apache, Camel, Cxf, Ejb, Gerrit, Git, Hadoop, Hdfs, Hibernate, I18n, Inetd, Java, Jaxb, Jdk, Jira, Jpa, Jsp, Jsps, Jvm, Launchd, Log4j, Mapreduce, Maven, Memcache, Openssl, Passwd, Pwd, Readline, Rmi, Servlet, Servlets, Solaris, Solr, Ssh, Symlink, Symlinks, Tomcat, Unix, Vim, Webapp, Webapps, Xinetd, and Yarn.
\ok{4813, Ant, Apache, Applet, Argouml, Auths, Bzip2, Camel, Columba, Command.com, Crlf, Cvs, Cxf, Distribution, Ebcdic, Ejb, Emf, File1, Foo.bar, Gerrit, Git, Hadoop, Hdfs, Hibernate, I18n, Inetd, Java, Jaxb, Jdk, Jedit, Jfreechart, Jira, Jmeter, Jpa, Jruby, Jsp, Jsps, Junit, Jvm, Jws, Kaffe, Launchd, Linux, Log4j, Mapreduce, Maven, Memcache, Myisam, Namespaced, Nls, Ocl, Openssl, Passwd, Pojo, Postgres, Prepending, Pwd, Readline, Rmi, Servlet, Servlets, Solaris, Solr, Squirrel, Ssh, Svn, Symlink, Symlinks, Tmp, Tomcat, Unix, Usecase, Utf, Vim, Webapp, Webapps, Xerces, Xinetd, and Yarn.}
% \todo{For \textit{@abstractproduct} are produce by find the similarity between each words in comments and project names in datasets which include Apache, Camel, Tomcat, Hadoop, Gerrit, and Log4j.
% Based on the figure~\ref{fig:similarity} the similarity drops drastically from 1.0 which we set  threshold at 0.9.
% Similarity are calculated by using tools namely Spacy~\cite{spacy2}.}
%\newline

We apply this process because
%In this step, we abstract parts of comment to their types by abstracting following types: URL, date, bug id, release version, product name.
%Each type abstracted into difference string (@abstracturl, @abstractdate, @abstractbugid, @abstractversion, @abstractproduct respectively).
we are more interested in the existence of these types rather than the actual terms, which do not appear frequently. 
For example, considering the comment ``\texttt{TODO: CAMEL-1475 should fix this}'', CAMEL-1475 will be changed to the string ``\texttt{@abstractproduct} \texttt{@abstractbugid}''.
Table~\ref{tab:reg} summarizes the regular expressions we used for identifying targeted terms.
Replacements using the regular expressions are conducted from top to bottom in the table. Subsequently, URLs linking to specific ids of bugs are abstracted to ``\texttt{@abstracturl @abstractbugid}''.
%we more interest that it is a product name follow by bug id more than word camel and number 1475.
%\subsubsection{Lemmatization}

%\noindent
\paragraph{Lemmatization.}

Lemmatization is a process to reduce the inflection form of words into dictionary form by considering the context in the sentences.
This process is applied to increase the frequency of words appearing by changing words into dictionary forms using tools from Spacy~\citep{spacy2}.

%\subsubsection{Special character removal}
%\noindent
\paragraph{Special character removal.}
Since non-English characters and non-numeric ones do not represent words, we use the regular expression \texttt{{[}\textasciicircum A-Za-z0-9{]}+} to remove them.
Stop word removal is not applied in this work because a stop word list contains important keywords for identifying ``on-hold'' self-admitted technical debt (e.g., when, until).
\ok{We use Spacy to apply lemmatization which will change words into their dictionary form. However, some single characters will appear, e.g., when lemmatising ``// TODO: Removed from UML 2.x'' to ``todo remove from uml 2 x''.}

\paragraph{\ok{Feature selection.}}
\ok{Auto-sklearn includes two feature selection functions from the sklearn library, sklearn.feature\textunderscore selection.GenericUnivariateSelect (Univariate feature selector) and  sklearn.feature\textunderscore selection.SelectPercentile (Select features according to percentile).}
\ok{Calling these functions is part of Auto-sklearn's feature preprocessing---it selects suitable feature processing based on meta-learning automatically.
}

% \subsection{Feature Extraction using N-gram IDF}
\subsection{N-gram Feature Extraction}
%From word abstraction,
% \subsubsection{Feature Extraction using N-gram IDF}
We extract n-gram term features by applying N-gram IDF~\citep{Shirakawa:2015,Shirakawa:2017:IWN:3077622.3052775}.
Inverse Document Frequency (IDF) has been widely used in many applications because of its simplicity and robustness; however, IDF cannot handle phrases (i.e., groups of more than one term). Because IDF gives more weight to terms occurring in fewer documents, rare phrases are assigned more weight than good phrases that would be useful in text classification.
N-gram IDF is a theoretical extension of IDF for handing multiple terms and phrases by bridging the theoretical gap between term weighting and multi-word expression extraction~\citep{Shirakawa:2015,Shirakawa:2017:IWN:3077622.3052775}.

%In software engineering research area, 
\cite{8094457}~reported that for classifying bug reports into bugs or non-bugs, classification models using features from N-gram IDF outperform models using topic modeling features.
In addition to this, we consider that n-gram word features are beneficial for comment classification rather than topic modeling because source code comments are generally short and contain only a small number of words.

\cite{8661216}~created classification models to identify design and requirement self-admitted technical debt using source code comments. By using N-gram IDF and auto-sklearn automated machine learning, classification models outperform models with single word features.

%After applying N-gram IDF on abstract comments, we extract dominant N-grams of any length from comment. 
In this study, we use an N-gram Weighting Scheme tool~\citep{GitHubiw51:online},
which uses an enhanced suffix array~\citep{Abouelhoda:2004:RST:985384.985389} to enumerate valid n-grams.
%which capable of handling terms of any length that $>=$ 1.
%The result of this step is features that contain all of the valid N-gram.
We obtain a list of all valid n-gram terms that contain at most 10-gram terms in the self-admitted technical debt comments and remove n-grams which have frequencies equal to one.
%Although previous work applied feature selection methods to decrease the number of n-gram terms used in classification models~\citep{8094457}, we use all n-gram terms in this study. Compared to more than ten thousand n-gram terms derived from bug reports~\citep{8094457}, 
We obtained about five thousand n-gram terms from our self-admitted technical debt comments.

\subsection{Classifier Learning}
\label{sec:cl}

Given the set of n-gram term features from the previous step,
we build a classifier that can identify ``on-hold'' self-admitted technical debt by classifying self-admitted technical debt comments into ``on-hold'' or not.

In machine learning, two problems are known: (1) no single machine learning method performs best on all data sets, and (2) some machine learning methods rely heavily on hyperparameter optimization.
%Classifier contains two classes (i.e.,  SATD with condition and SATD without condition).
% We prepare two classifiers for our evaluation, namely, random forests and automated machine learning.
% We use the random forest classifiers to compare the performance of n-gram term features with one-term features (as a benchmark), and to investigate important features for both classifiers.
%We prepare classifier for our evaluation, namely, automated machine learning.
%We first find important features by applying random forest. 
Automated machine learning aims to optimize choosing a good algorithm and feature preprocessing steps~\citep{NIPS2015_5872}. To obtain the best performance (RQ2.1), similar to~\cite{8661216}'s work, we apply auto-sklearn~\citep{NIPS2015_5872}, a tool of automated machine learning. %which also applies feature selection.

 Auto-sklearn addresses these problems as a joint optimization problem~\citep{NIPS2015_5872}. Auto-sklearn includes 15 base classification algorithms, and produces results from an ensemble of classifiers derived by Bayesian optimization~\citep{NIPS2015_5872}.

For classifier learning, we prepare feature vectors with N-gram TF-IDF scores of all n-gram terms. The score is calculated with the following formula:
%We then use automated-machine learning (i.e., auto-sklearn \cite{NIPS2015_5872}) to classify comments
%by applying score for each feature based on equation*~(\ref{eq:vector_score}) which extend from N-gram IDF weight1 formula\cite{GitHubiw51:online}:
\begin{equation*}\label{eq:vector_score}
     \textit{n-gram TF-IDF} = log(\frac{|D|}{sdf}) * gtf
\end{equation*}
where $|D|$ is the total number of comments, 
$sdf$ is the document frequency of a set of terms composing an n-gram, and
$gtf$ is the global term frequency.

\subsection{On-hold Condition Detection}

\ok{After ``on-hold'' self-admitted technical debt comments are identified, we try to identify their ``on-hold'' conditions. During our annotation, we found conditions of self-admitted technical debt that are related to waiting for a bug to be fixed, a release of a library, or a new version of a library.}

\begin{itemize}
    \item \ok{For a bug to be fixed, we abstract the bug report number. In a bug report tracking system, the bug report number is created by using the project name and report number which we abstract using the keywords @abstractproduct and @abstractbugid.}
    \item \ok{For release date, we abstract it using the keyword @abstractdate.}
    \item \ok{For a new version of a library, the library version usually appears in a project name and release version (e.g., 1.9.3, 4.0), which we abstract using the keywords @abstractproduct and @abstractversion.}
    \end{itemize}
As we have already replaced these terms with specific keywords shown in Table~\ref{tab:reg}, we can derive conditions by recovering the original terms.
The following is our detection process.
\begin{enumerate}
    \item Extract keywords of \texttt{@abstractdate}, \texttt{@abstractversion}, \texttt{@abstractbugid}, and \texttt{@abstractproduct} by preserving the order of appearance in the identified ``on-hold'' self-admitted technical debt comments.
    \item Group keywords to make valid conditions. Only the following sets of keywords are considered to be valid conditions, and other keywords that do not match the following orders are ignored.
    \begin{itemize}
        \item \texttt{\{@abstractdate\}}: an individual date expression.
        \item \texttt{\{@abstractproduct, @abstractversion, ...\}}: a product name followed by one or more version expressions, to indicate specific versions of the product.
        \item \texttt{\{@abstractproduct, @abstractbugid, ...\}}: a product name followed by one or more bug ID expressions, to indicate specific bugs of the product.
    \end{itemize}
\end{enumerate}

Identifying these keywords as conditions is not trivial, because they also frequently appear in comments that do not indicate ``on-hold'' self-admitted technical debt.
Since we limit this detection to the identified ``on-hold'' comments, we expect that this simple process can work. %The results for this step are described in Section~\ref{sec:rq22}.

\section{Classifier Evaluation}
\label{sec:evaluation}

In this section, we describe the steps we took to evaluate our classifier.

\subsection{Data Preparation and Annotation}

\begin{table}
    \centering
    \caption{Annotated self-admitted technical debt comments}
    \begin{tabular}{llr}
    \toprule
    & \textbf{characteristic} & \textbf{number} \\
    \midrule
    % \multirow
    % {2}{*}{excluded} & duplication & 92 \\
    \ok{excluded} & not self-admitted technical debt & 277 \\
    \midrule
    \multirow{2}{*}{classification data} & with condition (on-hold) & \ok{293} \\
    & without condition & \ok{5,236} \\
    \midrule
    sum & & \ok{5,806} \\
    \bottomrule
    \end{tabular}
    \label{tab:annotation}
\end{table}

\begin{table}[]
    \caption{Number of ``on-hold'' self-admitted technical debt comments in each project}
    \centering
    \begin{tabular}{lrl}
        \toprule
        \textbf{project} & \textbf{number} & \textbf{example} \\
        \midrule
        Apache Camel & 104 & // @deprecated will be removed on Camel 2.0 ... \\
        Apache Tomcat & 27 & // TODO This can be fixed in Java 6 ... \\
        Apache Hadoop & 23 & // TODO need to get the real port number \\ & & MAPREDUCE-2666 \\
        Gerrit Code Review & 6 & // TODO: remove this code when Guice fixes\\ & & its issue 745 \\
        Apache Log4j & 1 & // TODO: this method should be removed if \\ & & OptionConverter becomes a static \\
        \ok{Apache Ant} & \ok{7} & \ok{// since Java 1.4 ...} \\ 
        & & \ok{// workaround for Java 1.2-1.3}  \\
        \ok{Apache Jmeter} & \ok{2} & \ok{// TODO this bit of code needs to be tidied up} \\
        & & \ok{... Bug 47165} \\
        \ok{Argouml} & \ok{77} & \ok{// TODO}: \ok{gone in UML 2.1} \\
        \ok{Columba} & \ok{0} & \ok{--} \\
        \ok{EMF} & \ok{1} & \ok{// Note: Registry based authority is being} \\
                            & & \ok{removed ... which would obsolete RFC 2396.} \\
                            & & \ok{If the spec is added ... needs to be removed.} \\
        \ok{Hibernate} & \ok{5} & \ok{// FIXME Hacky workaround to JBCACHE-1202} \\
        \ok{JEdit} & \ok{6} & \ok{// undocumented hack to allow browser} \\
                                & & \ok{actions to work. // XXX - clean up in 4.3} \\
        \ok{JFreeChart} & \ok{2} & \ok{// TODO: In JFreeChart 1.2.0 ...} \\
        \ok{JRuby} & \ok{23} & \ok{// Workaround for JRUBY-4149} \\
        \ok{SQuirrel} & \ok{9} & \ok{// We know this fails - Bug\# 1700093} \\
        \midrule
        total & \ok{293} & -- \\
        \bottomrule
    \end{tabular}
\label{tab:numberonhold}
\end{table}

As shown in Figure~\ref{fig:condition}, we found fewer than 30 ``on-hold'' self-admitted technical debt comments in the sample of 335 comments. Since it is difficult to train classifiers on such a small number of instances, we investigated all \ok{2,507} comments again to prepare data for our classification. %Among all 2,599 comments, we found 92 duplicate comments (i.e., the exact same comment appeared in more than one location). After removing duplicate comments, 
\ok{After that, }the first and third author separately annotated the remaining comments in terms of (i) whether comments represent self-admitted technical debt (similar to Figure~\ref{fig:commentisdebt}) and (ii) whether the self-admitted technical debt comments include a condition (similar to Figure~\ref{fig:condition}). All conflicts in this annotation were resolved by the second author. Note that we decided to train the classifier on comments which had been removed through the resolution of self-admitted technical debt to ensure we were able to consider the entire lifecycle of the self-admitted technical comment before deciding whether to consider it on-hold.

\ok{
We also include a data set from ten open source projects introduced by~\cite{Maldonado2017}. First, we randomly selected a sample of 30 comments out of all 3,299 comments. The first author, third author, and an external annotator annotated these comments, resulting in 97.78\% overall agreement, i.e., ``almost perfect'' according to ~\cite{viera2005understanding}. Then the first author annotated the remaining comments.
}

\begin{comment}
Tables~\ref{tab:annotation} and~\ref{tab:numberonhold} show the result of this data preparation. From 
% 2,599 comments, 92 duplicate comments
\ok{2,507} and 277 comments that do not represent self-admitted technical debt are excluded. We obtained 161 ``on-hold'' comments and 2,069 other comments, which are used for our classification.
\end{comment}

Tables~\ref{tab:annotation} and~\ref{tab:numberonhold} show the result of this data preparation. From \ok{5,806} comments, 
% 92 duplicate comments and
277 comments that do not represent self-admitted technical debt are excluded. We obtained \ok{293} ``on-hold'' comments and \ok{5,236} other comments, which are used for our classification.

\subsection{Evaluation Settings}

We measure the classification performance in terms of precision, recall, $F_1$, and AUC. 
AUC is the area under the receiver operating characteristic curve.
The receiver operating characteristic curve is created by plotting the true positive rate (TPR) against the false positive rate (FPR) at various threshold settings.

\begin{equation*}\label{eq:precision}
    Precision = \frac{tp}{tp+fp}
\end{equation*}
\begin{equation*}\label{eq:recall}
    Recall = \frac{tp}{tp+fn}
\end{equation*}
\begin{equation*}\label{eq:$F_1$}
    F_1 = \frac{2 \cdot ( precision \cdot recall )}{( precision + recall )}
\end{equation*}
\begin{equation*}
    TPR = \frac{tp}{tp + fn}
\end{equation*}
\begin{equation*}
    FPR = \frac{fp}{tn + fp}
\end{equation*}

where $tp$ is the number of true positives,
$tn$ is the number of true negatives,
$fp$ is the number of false positives, and
$fn$ is the number of false negatives.

\paragraph{Comparison.}
A Naive Baseline is created based on the assumption that it is also possible to find ``on-hold'' technical debt comments while using basic searching similar to the \texttt{grep} command.
The words we use for searching are selected from the top 30 words that appear frequently in comments.
We manually classify words to select those that relate to ``on-hold'' technical debt.
The words we selected are \textit{``should''}, \textit{``when''}, \textit{``once''}, \textit{``remove''}, \textit{``workaround''}, \textit{``fixed''}, \textit{``after''}, and \textit{``will''}.

To assess the effectiveness of n-gram features in classifying ``on-hold'' self-admitted technical debt comments, we compare the performances of classifiers using N-gram TF-IDF and traditional TF-IDF~\citep{Salton:1988}. Except for feature extraction, the two classifiers are prepared using the same settings including term abstraction.

\paragraph{Ten-fold cross-validation.}
Ten-fold cross-validation divides the data into ten sets and every set is used as test set once while the others are used for training.
Due to the imbalance between the number of positive and negative instances, we use the Stratified ShuffleSplit cross validator\footnote{\url{http://scikit-learn.org/stable/modules/generated/sklearn.model_selection.StratifiedShuffleSplit.html}} of scikit-learn made available by~\cite{scikit-learn}, which intends to preserve the percentage of samples from each class. Because of this process, some instances can appear multiple times in different sets. Therefore we report the mean values of the evaluation metrics across all ten runs as the performance.

\ok{
To measure the effect of rebalancing on our classification, we compare the performance of N-gram TF-IDF with and without Stratified ShuffleSplit.
}

\subsection{RQ2.1 What is the best performance of a classifier to automatically identify ``on-hold'' self-admitted technical debt?}
\label{sec:rq21}

%We use the automated machine learning tool auto-sklearn~\citep{NIPS2015_5872} to assess the best performance a classifier can achieve when classifying on-hold self-admitted technical debt comments. 

\begin{table}
    \centering
    \caption{Performance comparison}
    \begin{tabular}{lrrrr}
    \toprule
     & & & & \ok{\textbf{N-gram TF-IDF}} \\
     & \textbf{naive baseline} & \textbf{TF-IDF} & \textbf{N-gram TF-IDF} & \ok{\textbf{without reblancing}} \\
    \midrule
    Precision & \ok{0.12} & \ok{0.60} & \ok{\textbf{0.63}} & \ok{0.59} \\
    Recall & \ok{0.64} & \ok{0.57} & \ok{\textbf{0.68}} & \ok{0.64} \\
    $F_1$ & \ok{0.20} & \ok{0.58} & \ok{\textbf{0.64}} & \ok{0.59} \\
    AUC & \ok{0.70} & \ok{0.78} & \ok{\textbf{0.83}} & \ok{0.80} \\
    \bottomrule
    \end{tabular}
    \label{tab:auto}
\end{table}

\begin{comment}
\begin{table}
    \centering
    \caption{Performance comparison}
    \begin{tabular}{lrrr}
    \toprule
     & \textbf{naive baseline} & \textbf{TF-IDF} & \textbf{N-gram TF-IDF} \\
    \midrule
    Precision & 0.19 & \textbf{0.79} & 0.72 \\
    Recall & \textbf{0.77} & 0.62 & 0.76 \\
    $F_1$ & 0.31 & 0.69 & \textbf{0.73} \\
    AUC & 0.76 & 0.81 & \textbf{0.87} \\
    \bottomrule
    \end{tabular}
    \label{tab:auto}
\end{table}
\end{comment}

\begin{table}[]
    \centering
    \caption{\ok{Cross-project classification on projects which contain on-hold more than 2\%}}
    \begin{tabular}{lrrrrr}
    \toprule
    \ok{\textbf{Project}} & \ok{\textbf{\% of on-hold}} & \ok{\textbf{Precision}} & \ok{\textbf{Recall}} & \ok{\textbf{F1}} & \ok{\textbf{AUC}} \\
    \midrule
    \ok{\textbf{Apache Ant}} & \ok{5.65\%} & \ok{0.33} & \ok{0.14} & \ok{0.20} & \ok{0.56} \\
    \ok{\textbf{Apache Camel}} & \ok{11.29\%} & \ok{0.54} & \ok{0.74} & \ok{0.63} & \ok{0.83} \\
    \ok{\textbf{Apache Hadoop}} & \ok{8.16\%} & \ok{0.61} & \ok{0.48} & \ok{0.54} & \ok{0.73} \\
    \ok{\textbf{Apache Tomcat}} & \ok{3.25\%} & \ok{0.16} & \ok{0.52} & \ok{0.25} & \ok{0.72} \\
    \ok{\textbf{ArgoUML}} & \ok{6.70\%} & \ok{0.45} & \ok{0.26} & \ok{0.33} & \ok{0.62} \\
    \ok{\textbf{Gerrit Code Review}} & \ok{5.45\%} & \ok{0.00} & \ok{0.00} & \ok{0.00} & \ok{0.44} \\
    \ok{\textbf{JEdit}} & \ok{2.55\%} & \ok{0.50} & \ok{0.67} & \ok{0.57} & \ok{0.82} \\
    \ok{\textbf{Jruby}} & \ok{4.98\%} & \ok{0.44} & \ok{0.35} & \ok{0.39} & \ok{0.66} \\
    \ok{\textbf{SQuirrel}} & \ok{3.95\%} & \ok{0.19} & \ok{0.44} & \ok{0.27} & \ok{0.68} \\ 
    \midrule
    \ok{\textbf{Average}} & - & \ok{0.36} & \ok{0.40} & \ok{0.35} & \ok{0.67} \\
    \bottomrule
    \end{tabular}
    \label{tab:cross}
\end{table}
%

% Since automated machine learning takes a lot of time and we achieved a better performance with N-gram TF-IDF compared to TF-IDF (see Section~\ref{sec:comparison}), we only evaluate the automated machine learning performance using N-gram TF-IDF.
% Because of the same time reason, we use ten-fold cross-validation instead of leave-one-out cross-validation.

%From the result of automated-machine learning,
As shown in Table~\ref{tab:auto},
our classifier with n-gram TF-IDF achieved a mean precision of \ok{0.63}, a mean recall of \ok{0.68}, a mean $F_1$-score of \ok{0.64}, and a mean AUC of \ok{0.83}. 
\ok{N-gram TF-IDF has the best performance in every evaluation.} % regarding F1 score.}
We consider that both precision and recall are essential for this kind of recommendation system.
Precision is important since false positives (i.e., unwarranted recommendations) will annoy developers. 
However, recall is still important since false negatives (i.e., recommendations that the system could have made but did not) might cause problems since developer will be unaware of important information.
\begin{comment}
our classifier with n-gram TF-IDF achieved a mean precision of 0.72, a mean recall of 0.76, a mean $F_1$-score of 0.73, and a mean AUC of 0.87. 
%The result from Table~\ref{tab:auto} shows that
%While TF-IDF has the best performance with regard to precision, n-gram TF-IDF has the best performance regarding F1 score and AUC. 
%We use automated machine learning with ten-fold cross-validation.
%Compared to TF-IDF classifier shown in Table~\ref{tab:auto} with higher precision.
% our classifier achieved a mean precision of 0.81, a mean recall of 0.67, and a mean $F_1$-score of 0.72. Compared to the previous result only with random forests shown in Table~\ref{tab:forest}, recall improved substantially with still high precision and a higher $F_1$-score.
We consider that both precision and recall are essential for this kind of recommendation system.
Precision is important since false positives (i.e., unwarranted recommendations) will annoy developers. 
However, recall is still important since false negatives (i.e., recommendations that the system could have made but did not) might cause problems since developer will be unaware of important information.
\end{comment}
\ok{
We also run an experiment for cross-project classification on projects for which the ratio of ``on-hold'' self-admitted technical debt comments among all self-admitted technical debt comments is more than 2\%. For cross-project classification, we divide data into sets according to their project. 
    Every set is used as test set once while the other sets are used for training.
    Table~\ref{tab:cross} shows the results for each project. On average, our classifier with cross-project classification achieved a mean precision of 0.36, a mean recall of 0.40, a mean $F_1$-score of 0.35, and a mean AUC of 0.67.
}

\subsection{RQ2.2 How well can \ok{our classifier} automatically identify the specific conditions in ``on-hold'' self-admitted technical debt?}
\label{sec:rq22}

\begin{table}[]
    \centering
    \caption{Examples of specific conditions in ``on-hold'' self-admitted technical debt comments}
    \begin{tabular}{ll}
        \toprule
        \textbf{specific condition} & \textbf{example of ``on-hold'' comments} \\
        \midrule
        @abstractdate & // Workaround for, Adobe Read 9 plug-in on \\
         & \ok{IE bug}  // Can be removed after \textbf{26 June 2013} \\
        @abstractproduct, @abstractversion & // TODO cmueller:, remove the  \\
        & ``httpBindingRef'' look up in \textbf{Camel 3.0} \\
        @abstractproduct, @abstractbugid & // FIXME (\textbf{CAMEL-3091}): @Test \\
        \bottomrule
    \end{tabular}
    \label{tab:spcondition}
\end{table}

Because of our treatment of imbalanced data (see Section~\ref{sec:rq21}), some comments can appear multiple times in the test set. We consider that an ``on-hold'' comment is correctly identified only if it has been classified correctly in all cases where it was part of the test set. 
% Previous version 121 on-hold comments
Our classifier was able to identify \ok{230 out of 293} ``on-hold'' comments correctly. 
% Previous version 66 comments
Among them, \ok{80} comments contain abstraction keywords which indicate a specific condition, and all those instances were confirmed to be specific conditions by manual investigation.
Some comments do not mention specific conditions, such as ``\texttt{This crap is required to work around a bug in hibernate}''.
%We could identify conditions in 55$\%$ of identified on-hold comments.
% Previous version: Among the 19 false positives (incorrectly identified comments), 12 comments contain abstraction keywords
Among the \ok{168} false positives (incorrectly identified comments), \ok{9} comments contain abstraction keywords, but these keywords are used for references and not for conditions that a developer is waiting for. 
% Previous version: In summary, 84$\%$ (66/(12+66)) of the detected specific conditions are correct, and for 55$\%$ (66/121) 
In summary, \ok{90$\%$ (80/(9+80))} of the detected specific conditions are correct, and for \ok{35$\%$ (80/230)} of the ``on-hold'' comments, we were able to identify the specific condition that a developer was waiting for.

% example for on-hold SATD

\section{Implications}
\label{sec:implications}

The ultimate goal of our work is to enable the automated management of \ok{certain kinds of} self-admitted technical debt. Previous work~\citep{Zampetti2018} has found that most changes which address self-admitted technical debt require complex code changes---as such, it is unrealistic to assume that automated tool support could handle all kinds of requirement debt and design debt that developers admit in source code comments. Thus, in this work we set out to first identify a sub-class of self-admitted technical debt amenable to automated management and second develop a classifier which can reliably identify this sub-class of debt.

Our qualitative study revealed one particular class of self-admitted technical debt potentially amendable to automated tooling: ``on-hold'' self-admitted technical debt, i.e., comments in which developers express that they are waiting for a certain external event or updated functionality from an external library before they can address the debt that is expressed in the comment. In other words, the comment is ``on hold'' until the condition has been met.

Based on the data set made available by~\cite{MaldonadoICSME2017} \ok{and~\cite{Maldonado2017}}, we identified a total of \ok{230} comments which indicate ``on-hold'' self-admitted technical debt, confirming that this phenomenon is prevalent and exists in different projects. Our classifier to identify ``on-hold'' self-admitted technical debt was able to reach an AUC of \ok{0.83} in identifying comments that belong to this sub-class. In addition, we were able to identify specific conditions contained within these comments (\ok{90}$\%$ of conditions are detected correctly).
\ok{    Based on 15 projects, there are 293 ``on-hold'' comments out of 5,529 self-admitted technical debt comments, resulting in a relative frequency of 5.30\%. Out of 15 projects, the ratio of ``on-hold'' comments compared to all self-admitted technical debt comments is larger than 2\% for nine projects.
}
%\todo{details}.

\begin{figure*}
\centering
\begin{lstlisting}[language=Java]
  /*
  * TODO: After YARN-2 is committed, we should call containerResource.getCpus()
  * (or equivalent) to multiply the weight by the number of requested cpus.
  */
\end{lstlisting}
\caption{On-hold self-admitted technical debt example which we correctly identify, cf.~\url{https://github.com/apache/hadoop/commit/80eb92aff02cc9f899a6897e9cbc2bc69bd56136/hadoop-yarn-project/hadoop-yarn/hadoop-yarn-server/hadoop-yarn-server-nodemanager/src/main/java/org/apache/hadoop/yarn/server/nodemanager/util/CgroupsLCEResourcesHandler.java}}
\label{fig:onholdcorrectsample}
\end{figure*}

\begin{figure*}
\centering
\begin{lstlisting}[language=Java]
 /**
  * Ugly workaround because CodeMirror never hides lines completely.
  * TODO: Change to use CodeMirror's official workaround after
  * updating the library to latest HEAD.
  */
\end{lstlisting}
\caption{On-hold self-admitted technical debt example which our classifier cannot identify, cf.~\url{https://github.com/gerrit-review/gerrit/commit/0485172aaa70e3b1f0e98c00215672657e6f462e/gerrit-gwtui/src/main/java/com/google/gerrit/client/diff/CodeMirrorDemo.java}}
\label{fig:onholdincorrectsample}
\end{figure*}

\begin{table}[]
    \centering
    \caption{Top 10 N-gram TF-IDF frequent features only appear in on-hold comments.}
    \begin{tabular}{lr}
        \toprule
        \ok{\textbf{N-gram Features}} & \ok{\textbf{frequency}}\\
        \midrule
        \ok{`remove', `in', `abstractproduct', `abstractversion'} & \ok{7} \\
        \ok{`in', `uml', '2', `x'} & \ok{7} \\
        \ok{`fix', `in'} & \ok{7} \\
        \ok{`workaround', `to'} & \ok{6} \\
        \ok{`todo', `cmueller', `remove', `the'} & \ok{6} \\
        \ok{`ref', `attribute'} & \ok{6} \\
        \ok{`be', `remove', `in', `abstractproduct', `abstractversion'} & \ok{6} \\
        \ok{`workaround', `for', `abstractproduct', `abstractbugid'} & \ok{5} \\
        \ok{`for', `abstractversion'} & \ok{5} \\
        \ok{`after', `abstractproduct', `abstractbugid'} & \ok{5} \\
        \bottomrule
    \end{tabular}
    \label{tab:ngram_feature}
\end{table}

\ok{Table~\ref{tab:ngram_feature} shows the top features from N-gram TF-IDF ranked by how frequently our classifier uses them to distinguish ``on-hold'' comments from other self-admitted technical debt comments.
    Figure~\ref{fig:onholdcorrectsample} shows an example of an ``on-hold'' self-admitted technical debt comment. Our model can identify conditions using the keywords @abstractproduct and @abstractbugid referring to YARN-2.
    Figure~\ref{fig:onholdincorrectsample} shows an example that our classifier could not identify correctly. The ``on-hold'' condition refers to a workaround waiting for an update to the CodeMirror library.
}

Given all the events and new releases that happen in a software project at any given point in time, it is unrealistic to assume that developers will be able to stay on top of all instances of technical debt that are ready to be addressed once a condition has been met. Instead, there is a risk that developers forget to go back to these comments and debt instances even when the event they were originally waiting for has occurred.

This work builds a first step towards the design of automated tools that can support developers in addressing \ok{certain kinds of} self-admitted technical debt. In particular, based on the classifier introduced in this work, it is now possible to build tool support which can monitor the specific external events we have identified in this work (e.g., certain bug fixes or the release of new versions of external libraries) and notify developers as soon as a particular debt is ready to be addressed. \ok{While the ratio of ``on-hold'' comments is fairly low, such comments appeared in almost all of the studied projects, and we argue that alerting developers when such comments are ready to be addressed can prevent bugs or vulnerabilities that might otherwise occur, e.g., because of outdated libraries.}

\ok{In terms of tool support, we envision a tool which supports the developer by indicating comments that are ready to be addressed rather than a tool which addresses comments automatically. Addressing comments automatically---even though it is an interesting research challenge---is problematic for two reasons: (1) the precision of such a tool would have to be really high, and current work including our own suggests that this is not yet the case; and (2) developers are unlikely to relinquish control over their code base to a tool which automatically changes code.} 

\section{Threats to Validity}
\label{sec:threats}

% In terms of \textit{construct validity}, we use precision, recall and $F_1$-score as our evaluation metrics, similar to previous work that requires classification~\citep{MaldonadoICSME2017,8094457}. 
% Due to the limitation of time and computation, we cannot provide a comparison between N-gram TF-IDF and traditional TF-IDF on automated machine learning. 
% However, our comparison between N-gram TF-IDF and traditional TF-IDF using random forests showed that N-gram TF-IDF outperforms traditional TF-IDF.

Regarding threats to \textit{internal validity}, it is possible that we introduced bias through our manual annotation. While we generally achieved high agreement regarding the annotation questions listed in Table~\ref{tab:annotationquestions}, the initial agreement regarding RQ1.1 was low which is explained by the nature of the open-ended question. We resolved all disagreements through multiple co-located coding sessions with the first three authors of this paper. Note that we do not use the results of RQ1.1 as an input for our classifier. We may possibly have wrongly classified the removal of self-admitted technical debt, since in particular for comments indicating the need for new features, it can be hard to judge whether the new feature was indeed fully implemented.

For \textit{external validity}, while we analyzed a statistically representative sample of commits for RQ1 and the entire data set made available by~\cite{MaldonadoICSME2017} (after removing duplicates) for RQ2, we cannot claim generalizablity beyond the projects contained in this data set and our classifier might be biased as a result of the small number of projects. The limited data set allowed us to perform an in-depth qualitative analysis, and future work will need to investigate the applicability of our results to other projects and within-project prediction.

\ok{
For \textit{construct validity}, this is related to the manual labeling of ``on-hold'' self-admitted technical debt. A label might be affected by annotator misunderstand or mislabeling. Despite annotators resolving disagreements through discussion, the labels might still be incorrect.}

\section{Related Work}
\label{sec:related}

%Related work on self-admitted technical debt.

Self-admitted technical debt has been a popular research topic in the software engineering community in recent years. In this section, we introduce key research related to our study.

\subsection{\ok{Impact of self-admitted technical debt}}

\cite{surveysatd} conducted a survey about self-admitted technical debt by investigating three categories: (i) detection, (ii) comprehension, and (iii) repayment.
Detection focuses on identifying and detecting self-admitted technical debt.
Comprehension studies the life cycle of self-admitted technical debt. 
Repayment focuses on removal of self-admitted technical debt. 
This research found a lack of research related to the repayment of self-admitted technical debt.

\cite{MaldonadoICSME2017} studied the removal of self-admitted technical debt by applying natural language processing to self-admitted technical debt. They found that (i) the majority of self-admitted technical debt was removed, (ii) self-admitted technical debt was often removed by the person who introduced it, and (iii) self-admitted technical debt lasts between 18 to 172 days (median). Using a survey, the authors also found that developers mostly use self-admitted technical debt to track bugs and code that requires improvement. Developers mostly remove self-admitted technical debt when they are fixing bugs or adding new features.

\cite{Zampetti2018} conducted an in-depth quantitative and qualitative study of self-admitted technical debt. They found that (i) 20\% to 50\% of the corresponding comments were accidentally removed when entire methods or classes were dropped, (ii) 8\% of self-admitted technical debt removals were indicated in the commit messages, and (iii) most of the self-admitted technical debt requires complex changes, often changing method calls or conditionals.

\cite{7832911} introduced a large-scale empirical study across 159 software projects. From this data they performed manual analysis of 366 comments, showing (i) an average of 51 self-admitted technical debt comments per system, (ii) that self-admitted technical debt consists of 30\% code debt, 20\% defect debt, and 20\% requirement debt, (iii) the number of self-admitted technical debt comments is increasing over time, and (iv) on average it takes over 1,000 commits before self-admitted technical debt is fixed.

\cite{7476641} studied the relation between self-admitted technical debt and software quality based on five open source projects (i.e., Hadoop, Chromium, Cassandra, Spark, and Tomcat). 
Their result showed that (i) there is no clear evidence that files with self-admitted technical debt had more defects than other files, (ii) compared with self-admitted technical debt changes, non-debt changes had a higher chance of introducing other debt, but (iii) changes related to self-admitted technical debt were more difficult to achieve.

\cite{Mensah:2018} introduced a prioritization scheme.
After running this scheme on four open source projects, they found four causes of self-admitted technical debt which was code smells (23.2\%), complicated and complex task (22.0\%), inadequate code testing (21.2\%), and unexpected code performance (17.4\%). The result also showed that self-admitted technical design debt was prone to software bugs, and that for highly prioritized self-admitted technical debt tasks, more than ten lines of code were required to address the debt.

\cite{Kamei2016} used analytics to quantify the interest of self-admitted technical debt to see how much of the technical debt incurs positive interest, i.e., debt that indeed costs more to pay off in the future. They found that approximately 42--44\% of the technical debt in their case study incurred positive interest.

\cite{Palomba2017} conducted an exploratory study on the relationship between changes and refactoring and found that developers tend to apply a higher number of refactoring operations aimed at improving maintainability and comprehensibility of the source code when fixing bugs. In contrast, when new features are implemented, more complex refactoring operations are performed to improve code cohesion. In most cases, the underlying reasons behind the application of such refactoring operations were the presence of duplicate code or previously introduced self-admitted technical debt.

\cite{ddc2ab5db763415a8f1985c5a35b77a1}~propose a new technique to estimate Rework Effort, i.e., the effort involved to resolve self-admitted technical debt. They performed an exploratory study using text mining to extract self-admitted technical debt from source code comments. In order to extract source code comments, the authors apply text mining on four open source projects. The result from four projects shows a rework effort between 13 and 32 commented lines of code on average per self-admitted technical debt comment.

\subsection{\ok{Self-admitted technical debt identification and Classification}}

\cite{6976075} tried to identify self-admitted technical debt by looking into source-code comments in four open source project (i.e., Eclipse, Chromium OS, Apache HTTP Server, and ArgoUML).
Their study showed that (i) the amount of debt in these project ranged between 2.4\% and 31\% of all files, (ii) debt was created mostly by developers with more experience, and time pressures and code complexity did not correlate with the amount of self-admitted technical debt, and (iii) only 26.3\% to 63.5\% of self-admitted technical debt comments were removed.

\ok{\cite{Farias2015}~proposed a tool called CVM-TD (Contextualized Vocabulary Model for identifying Technical Debt) to identify technical debt by analyzing code comments. The authors performed an exploratory study on two open source projects. The result indicated that (1) developers use dimensions of CVM-TD when writing code comments, (2) CVM-TD provides vocabulary that may be used to detect technical debt, and (3) models need to be calibrated.
}

\ok{\cite{Farias2016} investigated the use of CVM-TD with the purpose of characterizing factors that affect the accuracy of the identification of technical debt, and the most chosen patterns by participants as decisive to indicate technical debt items. The authors conducted a controlled experiment to evaluate CVM-TD, considering factors such as English skills and experience of developers.}

\ok{\cite{silva2016does}~investigated the identification of technical debt in pull requests. The authors found that the most common technical debt categories are design, test, and project convention.}

% There are several studies classifying self-admitted technical debt comments.
\cite{Maldonado2017}~tried identifying design-related and requirement-related self-admitted technical debt using a maximum entropy classifier.

\cite{Huang:2018:IST:3188697.3188709}~tried classifying comments in terms of whether they contained self-admitted technical debt or not, and reported that their proposal outperformed the baseline method. 
% Since these studies used 1-gram terms as features, our proposal of using n-gram term features may improve the above classification performances.

\cite{7332619}~studied types of self-admitted technical debt using source code comments. This study classified types of self-admitted technical debt into design debt, defect debt, documentation debt, requirement debt, and test debt. The most common type of self-admitted technical debt is design debt and the second most common type is requirement debt. Self-admitted technical debt consist of 42\% to 84\% design debt, and 5\% to 45\% requirement debt.

\cite{Zampetti2017} developed a machine learning approach to recommend when design technical debt should be self-admitted. They found their approach to achieve an average precision of about 50\% and a recall of 52\%. When predicting cross-projects, the performance of the approach improved to an average precision of 67\% and a recall of 55\%. 

\cite{8352718}~identify self-admitted technical debt using change-level self-admitted technical debt determination. This model identifies whether a change introduces self-admitted technical debt. In order to create the model, they identified technical debt using all versions of source code comments. Then, they manually label changes that introduce technical debt in comments and extract 25 features which belong to three groups, i.e., diffusion, history, and message. After that, they create a classifier using random forest. Across seven projects, this model achieves an AUC of 0.82 and cost-effectiveness of 0.80.

\ok{\cite{Flisar2019}~developed a new method to detect self-admitted technical debt using word embedding trained from unlabeled code comments. They then apply feature selection methods (Chi-square, Information Gain, and Mutual Information), and use three classification algorithms (Naive Bayes, Support Vector Machine, and Maximum Entropy) to test on ten open source projects. Their proposed method was able to achieve 82\% correct predictions.}

\ok{\cite{Liu:2018:SDT:3183440.3183478} proposed a self-admitted technical debt detector tool which is able to detect debt comments using text mining and is able to manage detected comments in an IDE via an Eclipse plug-in.
}

\ok{\cite{Ren:2019:NND:3343019.3324916}~proposed a Convolutional Neural Network for classifying code comments as self-admitted technical debt or not, based on ten open source projects. Their approach outperforms text-mining-based methods both in terms of within-project and cross-project prediction.
}

In our study we use the same data set as previous research~\citep{MaldonadoICSME2017,Maldonado2017}.

\section{Conclusions and Future Work}
\label{sec:conclusions}

Self-admitted technical debt refers to situations in which software developers explicitly admit to introducing technical debt in source code comments, arguably to make sure that this debt is not forgotten and that somebody will be able to go back later to address this debt. In this work, we hypothesize that it is possible to develop automated techniques to manage a subset of self-admitted technical debt.

As a first step towards automating a part of the management of \ok{certain kinds of} self-admitted technical debt, in this paper, we contribute (i) a qualitative study on the removal of self-admitted technical debt in which we annotated a statistically representative sample of 335 technical debt comments using seven questions that emerged as part of the qualitative analysis; (ii) the definition of ``on-hold'' self-admitted technical debt (debt which contains a condition to indicate that a developer is waiting for a certain event or an updated functionality having been implemented elsewhere) which emerged from this qualitative analysis as a particular class of self-admitted technical debt that can potentially be managed automatically; and (iii) the design and evaluation of a classifier for self-admitted technical debt which can detect ``on-hold'' debt with an AUC of \ok{0.83} as well as identify the specific conditions that developers are waiting for.

Building on these contributions, in our future work we intend to build the tool support that our classifier enables: a recommender system which can indicate for a subset of self-admitted technical debt in a project when it is ready to be addressed. We found that self-admitted technical debt is sometimes addressed by uncommenting source code that has already been written in anticipation of the debt removal. As another step towards the automation of technical debt removal, in future work, we will explore whether it is possible to address such debt automatically.

%\begin{acknowledgements}
%\end{acknowledgements}

\bibliographystyle{spbasic}      % basic style, author-year citations

\end{sloppy}
\end{document}